\documentclass[notitlepage]{revtex4-1}

\usepackage{amsmath,amssymb}
\usepackage[bookmarks=false,colorlinks=true]{hyperref}
\usepackage{color}
\usepackage{latexsym}
\def\DR{\rm I\kern-1.45pt\rm R}
\def\DC{\kern2pt {\hbox{\sqi I}}\kern-4.2pt\rm C}

\newcommand{\ba}{\begin{array}}
\newcommand{\ea}{\end{array}}
\newcommand{\be}{\begin{equation}}
\newcommand{\ee}{\end{equation}}
\newcommand{\bea}{\begin{eqnarray}}
\newcommand{\eea}{\end{eqnarray}}

\def\bz{{\bar z}}

\def\bu{{\bar u}}

\begin{document}

\title{Symmetries in superintegrable deformations of oscillator/Coulomb systems: "holomorphic factorization"}
\author{Tigran Hakobyan}
\email{tigran.hakobyan@ysu.am}
\author{Armen Nersessian}
\email{arnerses@ysu.am}
\affiliation{Yerevan State University, 1 Alex Manoogian St., Yerevan, 0025, Armenia}
\affiliation{Tomsk Polytechnic University, Lenin Ave. 30, 634050 Tomsk, Russia}
\author{Hovhannes Shmavonyan}
\email{shmavonyanhov@gmail.com}
\affiliation{Yerevan State University, 1 Alex Manoogian St., Yerevan, 0025, Armenia}
\begin{abstract}
We  propose a unified description for the constants of motion for
superintegrable deformations of the oscillator and Coulomb systems on
$N$-dimensional Euclidean space, sphere and hyperboloid. We  also consider the duality between these generalized systems and present some example.
\end{abstract}
\maketitle
\section{Introduction}
The $N$-dimensional oscillator and Coulomb problem play special role among other integrable systems by many reasons.
One of the main reasons, due to which these models continue to attract permanent interest during the last centuries,
 is their "maximal superintegrability", i.e. the existence of maximally possible number, $2N-1$,  of functionally
 independent constants of motion. The
rational Calogero model with oscillator potential \cite{calogero} and its generalization associated with arbitrary Coxeter systems \cite{algebra},
 is superintegrable system as well \cite{woj83}.
 The oscillator and Coulomb systems admit obvious separation
of the radial and angular variables, which is useful to formulate in terms of conformal algebra $so(1,2)\equiv sl(2,\mathbb{R})$   defined by
the following Poisson bracket relations
\be
 \{  \mathcal{H}_0 , \mathcal{D}\}=2 \mathcal{H}_0  ,\qquad\{
\mathcal{H}_0 , \mathcal{K}\}=\mathcal{D},\qquad  \{ \mathcal{K},\mathcal{D}\}= -2\mathcal{K}.
\label{ca}
\ee
The generators $\mathcal{H}_0, \mathcal{K},\mathcal{D}$ could be  identified, respectively, with the Hamiltonian of some $N$-dimensional mechanical system,
and with the generators  of conformal boost and dilatation.This system is usually called "conformal mechanics", and $so(1,2)$ symmetry appears as its dynamical symmetry \cite{fubini}.
Introduce the effective "radius" and conjugated momentum,
 \be
r=\sqrt{2\mathcal{K}},\qquad p_r=\frac{\mathcal{D}}{\sqrt{2\mathcal{K}}},\qquad \{p_r, r\}=1,
\ee
and  define a Casimir of conformal algebra
\begin{equation}
{\cal I}=2{\cal H}_0{\cal K}-\frac12 {\cal D}^2:
\qquad
\{\mathcal{I},\mathcal{H}_0\}=\{\mathcal{I},\mathcal{K}\}=\{\mathcal{I},\mathcal{D}\}=0.
\label{ml}
\end{equation}
It  is obviously   a constant of motion  independent on radial coordinate and momentum, and thus
could be expressed via appropriate   angular coordinates $\phi_a $ and
canonically conjugate momenta $\pi_a $ which are independent on radial ones: ${\cal I}={\cal I}(\phi_a,\pi_a)$.
 In these terms the generators of conformal algebra read:
\begin{equation}
 {\cal H}_0=\frac{{p}^2_r}{2}+
\frac{{\cal I}}{r^2},
\qquad
 {\cal D}=r p_r ,
\qquad
{\cal K}=\frac{r^2}{2}.
%\qquad
% {\cal I}\equiv \frac{{\bf J}^2}{2}+  r^2V({\bf x}),
\label{so2}
\end{equation}
Hence, such a separation of angular and radial parts could be defined for any system with dynamical conformal symmetry, and for those with additional potentials be function of  conformal boost $\mathcal{K}$.
In particular, such a generalized oscillator and Coulomb systems assume adding of potential
\be
V_{osc}=\omega^2\mathcal{K},\qquad V_{Coul}=-\frac{\gamma}{\sqrt{2\mathcal{K}}},
\ee
 so that their Hamiltonian takes the form
 \be
\mathcal{H}_{osc/Coul}=\frac{p^2_r}{2}+\frac{\mathcal{I}}{r^2}+V_{osc/Coul}(r).
\label{planarHam}\ee
Well-known  generalizations  of oscillator and Coulomb systems to $N$-dimensional spheres and two-sheet hyperboloids (pseudospheres) \cite{higgs} can be described in a similar way (see Section 5).

 In Refs.~\cite{hlnsy,rapid} a separation of "radial" and "angular" variables has been used for constructing the integrable deformations of oscillator and Coulomb systems (and of their (pseudo)spherical generalizations) via replacement of the spherical part of pure oscillator/Coulomb
 Hamiltonians (quadratic casimir of $SO(N)$ algebra)  by some other integrable system
formulated  in terms of the action-angle variables.
%\be
%\mathcal{I}=\mathcal{I}(I_a),
%\qquad
%\Omega=\sum_{a=1}^{N-1} dI_a\wedge d\Phi_a.
%\ee
Analyzing these deformations in terms of action-angle variables, it was found that they are superintagrable iff
the spherical part has the form
\be
\mathcal{I}=\frac12\left(\sum_{a=1}^{N-1}k_a I_a + c_0 \right)^2%=\frac12 \left(\sum_a k_a u_a{\bar u}_a+c_0\right)^2
\label{angular0}
\ee
with $c_0$ be arbitrary constant and $k_a$ be rational numbers.
Moreover, it was demonstrated, by the use of the results of Ref.~\cite{flp}, that the angular part of rational Calogero model   belongs to this set of systems. Thus, it was concluded that rational Calogero model with Coulomb potential (Calogero-Coulomb system) is superintegrable system. Besides,  superintegrable generalizations of the rational Calogero models with oscillator/Coulomb potentials on the $N$-dimensional spheres and two-sheet hyperboloids have been suggested there. The explicit expressions of their symmetry generators and respective algebras have been given in
Refs.~\cite{runge,francisco}. An integrable  two-center  generalization  of the  Calogero-Coulomb systems (and those in the presence of Stark term, which was called Calogero-Coulomb-Stark model) has been also revealed \cite{Calogero-Stark}.
Other  superintegrable deformations of the two-dimensional oscillator and Coulomb systems of this kind  are
known as    Trembley-Turbiner-Wintenitz (TTW) system \cite{TTW} and Post-Winternitz (PW) system \cite{pw}.
They are defined by the Hamiltonians \eqref{planarHam} with the angular part  given by the P\"oschl-Teller system on circle
is a particular case of the Calogero-oscillator system, and their generalizations to sphere and hyperboloid \cite{hlnsy}
\be
\mathcal{I}_{PT}=\frac{p^2_\varphi}{2}+\frac{k^2\alpha^2}{\sin^2k\varphi}+\frac{k^2\beta^2}{\sin^2k\varphi},
\label{PT}\ee
where $k$ is an (half)integer.
The superintegrability  of these systems was observed initially by  numerical simulations, and only later    an  analytic expression for the
additional constant of motion was presented  \cite{kalnins}.
Initially these systems were invented as new superintegrable models but soon it was observed that
they  coincide  with the two-dimensional rational Calogero model with oscillator/Coulomb potential associated with the
dihedral group $D_k$ \cite{lny}. (Super)integrability of their (pseudo)spherical counterparts was noticed in Ref.~\cite{hlnsy}),
Nevertheless, their  study attracts much attention up to now. Among interesting observations in this subject was the so-called "holomorphic factorization" of constants of motions of these systems developed by M.Ran\"ada \cite{ranada}, which could be viewed as a classical counterpart of
the factorization of Schroedinger operator. In this approach all constants of motion of the two-dimensional oscillator/Coulomb systems were presented
 as a single complex integral, which was represented  as a product of two complex functions: one of the latter involves only "angular"  variables, and other- "radial"  ones and on \eqref{PT}. In our recent paper \cite{PANShmavon} it was observed that the radial part of this complex function is related with
coordinate parameterizing Klein model  Lobachevsky plane (so that the $so(1,2)$ generators
define  its Killing potentials (Hamiltonian generators of the
isometries of the K\"ahler structure), while  the angular part is related with angle variable of the P\"oschl-Teller Hamiltonian \eqref{PT}, in agreement  with Ref.~\cite{lobach}. This allowed to suggest the extension of  that construction to higher-dimensional (super)integrable systems with oscillator/Coulomb potential.

The goal of thjis paper is to present "holomorphic factorization" to the superintegrable generalizations of oscillator and Coulomb systems on $N$-dimensional Euclidean space, sphere and  two-sheet hyperboloid (pseudosphere).
For this purpose we parameterize the phase spaces of that system by the
 complex variable $z=p_r+\imath\sqrt{2\mathcal{I}}/r$
  identifying the radial phase subspace with the Klein model of
Lobachevsky plane  (compare with Refs.~\cite{lobach,PANShmavon}), and by the
 complex variables $u_a=\sqrt{I_a}{\rm e}^{\imath\Phi_a}$  unifying action-angle variables of the angular part of the systems.
We formulate, in these terms, the constants of motion of the systems under consideration and  calculate their  algebra.
Besides, we  extend to these systems the known oscillator-Coulomb duality transformation.

The paper is organized as follows:

In {\it Section 2} we introduce the appropriate complex coordinates unifying radial and angular variables and formulate the Poisson brackets and  generators of conformal algebra in these terms. Then we give "holomorphic factorization formulation" of the constants of motion of  higher-dimensional superintegrable conformal mechanics, and calculate their algebra.

In {\it Section 3} we formulate in these terms,  the higher-dimensional superintegrable generalizations of oscillator and Coulomb systems given by \eqref{so2},\eqref{angular0} and calculate the
algebra of their constants of motion.

In {\it Section 4} we  formulate, in this terms, the well-known oscillator-Coulomb duality transformation.

In {\it Section 5} we  extend the results of Section 2 to the systems  on $N$-dimensional sphere and two-sheet hyperboloid (pseudosphere).

Finally, in the {\it Section 6} we formulate in these terms, the special  of angular part of these systems.

\section{Conformal mechanics}
Let us consider the $N$-dimensional conformal mechanics, defined by the canonical symplectic structure ${d{\bf p} }\wedge {d{\bf x}}$ and the Hamiltonian
\begin{equation}
\label{h1}
\mathcal{H}_0=\frac{{\bf p}^2}{2}+V({\bf x})
\qquad {\rm with}
\qquad ({\bf x}\cdot\nabla) V({\bf x} )=-2V({\bf x}).
\end{equation}
The conformal algebra \eqref{ca} is generated  by the $\mathcal{H}_0$ and the generators of
 dilatation and conformal boost:
\begin{gather}
\mathcal{D}={\bf p}\cdot{\bf x},
\qquad  \mathcal{K}=\frac{{\bf x}^2}{2},
\label{gen}
\end{gather}
Extracting the radius $r=|{\bf x}|$ and its canonically conjugated momentum $\displaystyle p_r=\frac{{\bf p}\cdot{\bf x}}{r} $,  we can write
these  generators in the form \eqref{so2}
with ${\cal I}={\cal I}(\phi_a,\pi_a)$ be  the Casimir element  of the $so(1,2)$ algebra
and depending   on  the angular coordinates $\varphi_a $ and their
canonically conjugated momenta $\pi_a$.
Considered itself as a separate Hamiltonian, ${\cal I}$ describes
a particle on $(N-1)$-sphere, moving in the
field of the potential $U(\phi_a )=r^2V(\bf{x})$ (various aspects of these systems were studied in \cite{sphCal}).

To provide the conformal mechanics by integrability property we choose an integrable angular system, formulated
 in terms of the action-angle variables:
\be
\mathcal{I}=\mathcal{I}(I_a),
\qquad
\Omega=\sum_{a=1}^{N-1} dI_a\wedge d\Phi_a,\qquad \Phi_a\in [0,2\pi).
\ee
Introduce the complex variable $z$, identifying the radial phase subspace with the Klein model of
Lobachevsky plane  (compare with \cite{lobach,PANShmavon}),
 and complex variables $u_a$  unifying the action-angle variables:
\begin{equation}
{ z}=\frac{p_r}{\sqrt2}+\frac{\imath \sqrt{{\cal I}}}{r},
\qquad
u_a=\sqrt{I_a}{\rm e}^{\imath\Phi_a}\qquad{\rm with }\qquad {\rm Im}\,{ z}>0.
%,\qquad \Phi^a=\varphi^a+\frac{\Omega^a}{r^2}
 \label{px2}
\end{equation}
These variables have   the following nonvanishing  Poisson brackets:
\begin{equation}
\{{ z},\bar{ z}\}=-\frac{\imath ({ z}-\bar {z})^2}{2\sqrt{2{\cal I}}},
\qquad
\{u_a, {\bar u}_b\}=-\imath\delta_{ab},
\qquad
 \{z,u_a\}=-u_a\Omega_a\frac{\imath(\bar{z}-z)}{2\sqrt{2{\cal I}}},
 \qquad  \{z,{\bar u}_a\}= {\bar u}_{a}\Omega_a\frac{\imath(\bar{z}-z)}{2\sqrt{2{\cal I}}},
\label{z-z}
\end{equation}
where
\be
\Omega_a=\Omega_a(I)=\frac{\partial\sqrt{2\mathcal{I}}}{\partial I_a}.
\label{I}\ee
In these terms
 the generators of conformal algebra take the form
 \begin{equation}
\mathcal{H}_0=z\bar z,
\qquad \mathcal{D}= {\sqrt{2{\cal I}(u_a\bu_a)}}\frac{{ z}+\bar { z}}{\imath ({\bar z}- {
z})}, \qquad \mathcal{K}= \frac{{2{\cal I}(u_a\bu_a)}}{(\imath ({\bar z}- { z}))^2}.
\label{uk2}
\end{equation}
Note that  the action variables $I_a$ complemented with the Hamiltonian
form a set of Liouville integrals of the conformal mechanics \eqref{h1}. They
have a rather simple form while being expressed via the complex variables:
 \begin{gather}\label{Hzz}
\mathcal{H}_0=z\bar z,
\qquad
I_a = u_a\bar u_a:
\qquad 
\{H_0,I_a\}=\{I_a,I_b\}=0.
\end{gather}

Let us now look for the additional integrals of motion, if any.
%\be
%\Omega_a=\Omega_a(I) =\frac{\partial\sqrt{2\mathcal{I}}}{\partial I_a}.
%\ee
It is easy to verify using \eqref{z-z}, \eqref{Hzz}  that
\be
\{ze^{\imath \Lambda},\mathcal {H}_0\}=0
\qquad {\rm iff }\qquad
\{ \Lambda,\sqrt{2\mathcal{I} }\}=-1.
\ee
To get the single-valued function  we impose $\Lambda\in[0,2\pi )$ .
The   local  solutions of  the above equation  read
\be
\label{wa}
\Lambda_a=\frac{\Phi_a}{\Omega_a},
\ee
where $\Phi_a\in[0,2\pi)$ is angle variable and $I_a$ is given by \eqref{I}.
Therefore, the following local quantities are preserved and generate the set
of $N-1$ additional constants of motion:
\be
\label{Ma}
M_a=zu_a^\frac{1}{\Omega_a}=zI_a^\frac{1}{2\Omega_a}e^{\imath\frac{\Phi_a}{\Omega_a}},
\qquad
\{M_a,{\cal H}_0\}=0.
\ee
Using \eqref{px2}, \eqref{z-z}, one can verify that the only nontrivial  Poisson bracket
relations among them occur  between the conjugate $M_a$-s:
\begin{gather}
\label{M-M}
\qquad
\{M_a,M_b  \}=0,
\qquad
\{M_a,\overline{M}_b  \}
%=-\imath \delta_{ab}  \frac{ m_a^2}{I_a}{\cal M}_a\overline{\mathcal{ M}}_a
=-\frac{\imath\delta_{ab}}{\Omega_a^2} I_a^{\frac{1}{\Omega_a}-1}{\cal H}_0.
\end{gather}
However, for the generic $\Omega_a$, the constant \eqref{Ma} is not still globally well-defined,
since {$\Lambda\in [0,2\pi/\Omega_a)$.
To get the global solution for a certain coordinate $\Phi_a$,
we are forced to set $\Omega_a$ to a rational number:
\be
\label{ka}
\Omega_a=k_a=\frac{n_a}{m_a} ,\qquad m_a,n_a \in \mathbb{N}.
\ee
Then, taking $n_a$-th power for the locally defined conserved quantity, we get a globally defined
constant of motion for the system,
\be
\label{calM}
{\cal M}_a=M_a^{n_a}=z^{n_a}u_a^{m_a}=I_a^\frac{m_a}{2}z^{n_a}{\rm e}^{\imath m_a\Phi_a }.
\ee
Although both $M_a$ and ${\cal M}_a$ are complex, their absolute values
are expressed via Liouville integrals, and, hence,  do not produce new constants
of motion:
\be
\label{absM}
|M_a|^2=\mathcal{H}_0 I_a^\frac{1}{k_a},
\qquad
|{\cal M}_a|^2=\mathcal{H}_0^{n_a} I_a^{m_a}.
\ee
So, we have constructed  $2N-1$ functionally independent  constants of motion of the
generic superintegrable conformal mechanics \eqref{h1} with rational frequencies \eqref{wa}.
Therefore,  the conformal mechanics  will be superintegrable provided that the angular Hamiltonian has the form \eqref{angular0}
with  rational numbers $k_a$ \eqref{ka} and arbitrary constant $c_0$.\\

Full  symmetry algebra is given by the relations
\be
\label{M-bMcal}
\{{\cal M}_a,\overline{\mathcal{ M}}_b  \}
%=-\imath \delta_{ab}  \frac{ m_a^2}{I_a}{\cal M}_a\overline{\mathcal{ M}}_a
=-\imath \delta_{ab}  m_a^2 I_a^{m_a-1}{\cal H}_0^{n_a},\qquad \{H_0,{\cal M}_a\}=\{{\cal M}_a,{\cal M}_b  \}=0.
\ee
Note  that
\be
\{I_a,{\cal M}_b  \}=\imath \delta_{ab}M_b,\qquad \{H_0,I_a\}=\{I_a,I_b\}=0
\label{M-Mcal}
\ee
As we mentioned in Introduction, presented formulae are applicable not only for the nonrelativistic conformal mechanics on $N$-dimensional Euclidean space
defined by the Hamiltonian \eqref{h1} but for the generic finite-dimensional system with conformal symmetry, including relativistic one. Typical example of such a system is a particle moving in the  near-horizon limit of extreme black hole. Several examples of such systems were investigated by A.Galajinsky and his  collaborators (see Ref.~\cite{GalajBH} end refs therein).

\section{Deformed oscillator and Coulomb systems}
Let us extend the above consideration  to  the deformed $N$-dimensional oscillator and Coulomb systems defined by
the Hamiltonians
\be
\mathcal{H}_{osc/Coul}=\frac{p^2_r}{2}+\frac{\mathcal{I}}{r^2}+V_{osc/Coul}(r)
=z\bz+ V_{osc/Coul}(r),
\label{flat}\ee
where
\be
V_{osc}=\frac{ \omega^2 r^2}{2}=\omega^2\mathcal{K}=-\frac{{2 \omega^2{\cal I}}}{({\bar z}- { z})^2} ,
\qquad
V_{Coul}=-\frac{\gamma}{r}=-\frac{\gamma}{\sqrt{2\mathcal{K}}}=-\gamma\frac{\imath(\bz-z)}{2\sqrt{\mathcal{I}}}.
\ee
Clearly, the action variables of the angular mechanics $I_a$ together with the corresponding
Hamiltonian define  Liouville constants  of motion:
\be
\label{H-I}
\{H_{osc/Coul},I_a\}=\{I_a,I_b\}=0.
\ee
To endow  these systems by superintegrability  broperty we choose the angular part given by \eqref{angular0} with rational $k_a$, see \cite{rapid}.
Below we construct the additional constants of motion and calculate their algebra for both systems in terms of complex variables \eqref{px2}
introduced in previous section.

\subsubsection{Oscillator case}
The  $2N-2$  constants of motion of the deformed oscillator $\mathcal{H}_{osc}$ in the coordinates
\eqref{px2} are appeared to look as:
\be
\label{Mosc}
% {\cal M}^{osc}_a=\left(z^2-\Omega^2\frac{8\mathcal{ I} }{ ({\bar z}- { z})^2}
 {\cal M}^{osc}_a=\left(z^2-\frac{2\omega^2\mathcal{I}}{({\bar z}- { z})^2}\right)^{n_a}u_a^{2m_a},
\qquad
| {\cal M}_a^{osc}|^2=\left(\mathcal{H}_{osc}^2-2\omega^2\mathcal{I}\right)^{n_a}I_{a}^{2m_a}.
\ee
%with $
%r^2=-{{8 {\cal I}}}/{({\bar z}- { z})^2}.
%$.
The last equation together with \eqref{angular0} means that only the arguments of these complex quantities
give rise to new integrals independent of the Liouville ones.

In fact, they are based on the simpler quantities $A_a$ and $B_a$, which oscillate in time with the same frequency $w$:
\be
\label{Aa}
A_a=\left(z+\frac{\omega\sqrt{2\mathcal{I}}}{{\bar z}- { z}} \right)u_a^{\frac1{k_a}},
\quad
B_a=\left(z-\frac{\omega\sqrt{2\mathcal{I}}}{{\bar z}- { z}}\right)u_a^{\frac1{k_a}}:
\qquad\{{\cal H}_{osc}, A_a\}=\imath\omega  A_a,
\quad
\{{\cal H}_{osc}, B_a\}=-\imath\omega  B_a.
\ee
So, the product $A_a B_b$ is preserved,
\be
\label{H-AB}
\{{\cal H}_{osc}, A_a B_b\}=0,
\ee
but is not single valued.
Thus, we have to  take its $n_a$th power to get
a well defined constant of motion, which is precisely \eqref{Mosc}:
\be
\qquad
 {\cal M}^{osc}_a=(A_aB_a)^{n_a}.
\label{H-A}\ee
Note that the reflection $\omega\to -\omega$ in the parameter space maps between $ A_a$ and  $B_a$.
Together with complex conjugate, they are subjected to the following rules:
\be
\label{absAB}
| B_a |^2
 =\frac{{\cal H}_{osc}-\omega\sqrt{2{\cal I}}}{{\cal H}_{osc}+\omega\sqrt{2{\cal I}}} | A_a |^2,
 \qquad
 | A_a|^2=I_a^\frac{1}{k_a}\left({\cal H}_{osc}+\omega\sqrt{2{\cal I}}\right).
\ee
The complex observables  $ A_a$ and  $B_a$ are in involution,
\be
\label{A-A}
\{A_a, A_b\}=\{B_a, B_b\}=\{A_a, B_b\}=0,
\ee
so that the constants  of motion \eqref{Mosc} commute as well:
\be
\label{M-Mosc}
\{{\cal M}^{osc}_a,{\cal M}^{osc}_b  \}=0.
\ee
However, in contrast to the simplicity of the  relations \eqref{M-Mcal},
the Poisson brackets between ${\cal M}^{osc}_a$ and $\overline{\cal M}^{osc}_b$
are more elaborate. They can be derived from the Poisson brackets between
$ A_a$ and  $B_a$ and their conjugates having the following form:
\begin{gather}
\label{A-barB}
\{A_a,\bar B_b\}=-\frac{\imath \delta_{ab}}{k_a^2 I_a}A_a\bar B_a,
\qquad
\{\bar A_a, B_b\}=\frac{\imath \delta_{ab}}{k_a^2 I_a}\bar A_a B_a,
\\
\label{A-barA}
\{A_a,\bar A_b\}=
-\frac{ 2\imath\omega A_a\bar A_b }{\mathcal{H}_{osc}+\omega \sqrt{2{\cal I}}}
-\frac{\imath\delta_{ab}}{k_a^2}I_a^{\frac{1}{k_a}-1}
(\mathcal{H}_{osc}+\omega \sqrt{2{\cal I}}),
\\
\label{B-barB}
\{B_a,\bar B_b\}=
\frac{ 2\imath\omega A_a\bar A_b }{\mathcal{H}_{osc}-\omega \sqrt{2{\cal I}}}
-\frac{\imath\delta_{ab}}{k_a^2}I_a^{\frac{1}{k_a}-1}
(\mathcal{H}_{osc}-\omega \sqrt{2{\cal I}}).
\end{gather}
Hence, we have extended the "holomorphic factorization" formalism to the $N$-oscillator.
\subsubsection{Coulomb case}
The $2N-2$ locally defined integrals  of the generalized  Coulomb Hamiltonian can be written in the coordinates \eqref{px2}  as follows
 \be
 \label{Mcoul}
  M_a^{Coul}=\left(z-\frac{\imath\gamma}{2\sqrt{\cal{I}}} \right)u_a^{\frac{1}{k_a}},
  \qquad
 \{ {\cal H}_{Coul},   M_a^{Coul} \}=0.
 \ee
Like in  the previous cases,  only their arguments  produce  conserved quantities
independent from the Liouville integrals  \eqref{H-I}  since
 \be
 \label{absMc}
 \big| M_a^{Coul}\big|^2=\left(\mathcal{ H}_{Coul}+\frac{\gamma^2}{4\cal{I}}\right) I_{a}^\frac{1}{k_a}.
 \ee
They form the following
algebra, which can be verified using the Poisson brackets \eqref{z-z}:
 \be
 \label{M-Mcoul}
\big\{ M_a^{Coul},\overline{M}_b^{Coul}  \big\}
=
\frac{\imath\gamma^2 M_a^{Coul}\overline{M}_b^{Coul} }
    {\sqrt{{2\cal I}}(\gamma^2+4{\cal I}{\cal H}_{Coul})}
- \frac{\imath\delta_{ab}I_a^{\frac{1}{k_a}-1}}{k_a^2}\left({\cal H}_{Coul}+\frac{\gamma^2}{\sqrt{8{\cal I}}}\right), \qquad
  \big\{M_a^{Coul},M_b^{Coul}\big\}=0,
%\\
% \{{\cal M}_{(a) Coul},\mathcal{ H}_{Coul}\}=0,\quad \{{\cal M}_{(a) Coul},{\cal M}_{(b) Coul}  \}=0,
\ee
Let us also present the Poisson brackets of these quantities with Liouville constanst of motion
\be
\label{I-Mcoul}\big\{ I_a,M_b^{Coul}\big\}=\frac{\imath \delta_{ab}}{k_b}M_b^{Coul}.
\ee
Similar to the  previous cases, we are forced to take certain powers of the local quantities
\eqref{Mcoul} in order to get
the valid, globally defined  additional constants of motion of the deformed Coulomb problem:
\be
\label{cMcoul}
 {\cal M}_a^{Coul}=\big(M_a^{Coul}\big)^{n_a}=\left(z-\frac{\imath\gamma}{2\sqrt{\cal{I}}} \right)^{n_a}u_a^{ m_a}.
\ee
Their algebra can be deduced from the Poisson bracket relations \eqref{M-Mcoul} and \eqref{I-Mcoul}.
\\

So, in this Section we extended the method of "holomorphic factorization"  initially developed for the two-dimensional oscillator and Coulomb system, to the superintegrable generalizations of Coulomb and oscillator systems in any dimension.
For this purpose we parameterized the angular parts of these systems by action-angle variables.
To our surprise, we were able to get, in these general terms, the symmetry algebra of these systems.
Notice, that above formulae hold not only on the Euclidean spaces, but for the more general one, if we choose $\mathcal{I}$ be the system with a phase space different from $T_* S^{N-1}$.
\section{Oscillator-Coulomb correspondence}

As is known, the energy surface of the radial oscillator  can  be transformed to the energy surface  of the radial Coulomb problem by
 transformation ${\tilde r}=\lambda r^2$,${\tilde p}_{\tilde r}=p_r/2\lambda r$ where $r, p_r$ are radial coordinate and momentum of oscillator,
  ${\tilde r}, {\tilde p}_{\tilde r}$ are those of Coulomb problem, and $\lambda$ is     an arbitrary positive constant number (see,e.g.\cite{ter} for the review).
   Extension of oscillator-Coulomb correspondence  from the radial part  to the whole system, as well as to its quantum counterpart yields additional restrictions on the geometry   of configuration spaces.  Namely, only   $N=2,4,8,16$ -dimensional oscillator could be transformed
to the Coulomb system, that is $N=2,5,9$ dimensional Coulomb problem. These dimensions are distinguished due to Hopf maps  $S^1/S^0=S^1$,
 $S^3/S^1=S^2$, $S^7/S^3=S^4$, which allow to transform  spherical (angular) part of oscillator to those of  Coulomb probem.
  Indeed,  for the complete correspondence between oscillator end Coulomb system we should be able to transform the
  angular part of oscillator (that is particle on $S^{D-1}$)  to the angular part of Coulomb problem, i.e. to $S^{d-1}$.
  Thus, the only admissible dimensions  are $D=2,4,8,16$ and $d=2,3,5,9$.  In the first three cases we have to reduce
   the initial system by
   $Z_2$, $U(1)$ and $SU(2)$. For the latter case, in spite of many attempts, we do not know rigorous derivation of this correspondence,
 due  to the fact  that $S^7$ sphere has no Lie group structure. Respectively, in the generic case we  get the  extension of two-/three/five- dimensional Coulomb system specified by the presence $Z_2$/Dirac/$SU(2)$ Yang  monopole \cite{ntt}.
 In the  deformed Coulomb and oscillator problems  considered in this article  we do not require  that the angular parts of the systems should be spheres.
 Hence, trying to relate these systems we are not restricted by the  systems of mentioned dimensions.  Instead, we can try to relate the deformed oscillator and Coulomb systems  of the same dimension and find the restrictions to the structure of their angular parts.

Below we  describe this correspondence in terms complex variables introduced in previous Section.
Through this subsection  we will use "untilded" notation  for the  description of oscillator, and  the "tilded" notation for the description of Coulomb system.

The expression of the "Lobachevsky variable" \eqref{px2}
via radial coordinate and momentum  forces to relate the angular parts of oscillator and Coulomb problem by the expression $\tilde{\mathcal{I}}=\mathcal{I}/4$.
The latter  induces the following relations between "angle-like" variables $\Lambda,{\tilde\Lambda}$:  $\tilde{\Lambda}=2\Lambda$.
Alltogether  read
\be
{\tilde z}= \frac{\imath (\bz-z)}{\lambda\sqrt{\mathcal{I}} } z ,
\quad \tilde{\mathcal{I}}  = \frac{\mathcal{I}}{4},
\quad
\tilde{\Lambda}=2{\Lambda}
\qquad\Leftrightarrow\qquad
z=2\sqrt{\lambda} \sqrt[4]{\tilde{\mathcal{I}}} \frac{{\tilde z}}{\sqrt{\imath({\bar{\tilde z}}-{\tilde z})}},
\quad \mathcal{I}=4\tilde{\mathcal{I}},
\quad
 \Lambda=\frac{\tilde{\Lambda}}{2}.
\ee
This transformation is canonical in a sense, that preserve Poisson brackets between  $z,\bar z, \Lambda, \mathcal{I}$, and their tilded counterparts.
To make the transformation canonical, we preserve the angular variables unchanged $\tilde{u}_a=u_a$,  which implies to introduce for superintagrable systems the following identification
\be
\tilde{k}_a =\frac{k_a}{2}\qquad\Rightarrow \qquad \tilde{n}_a=n_a,\quad\tilde{m}_a=2m_a.
\ee
Then we can see, that this transformation relates the energy surfaces of oscillator and Coulomb systems:
\be
z\bz+\Omega^2\frac{{2{\cal I}}}{(\imath ({\bar z}- { z}))^2}-E_{\rm osc}=0
\qquad \Leftrightarrow \qquad
\frac{2\lambda \sqrt{\mathcal{\tilde{I}} } } {\imath(\bar{\tilde{ z} }- \tilde{z})}\left( \tilde{z}\tilde{\bz}-\gamma\frac{\imath(\tilde{\bz}-\tilde{z})}{2\sqrt{\mathcal{I}}}  - {\cal \tilde{E}}_{Coul}\right)=0,
\ee
where
\be
\tilde{\gamma}=\frac{E_{\rm osc}}{\lambda},\qquad \mathcal{\tilde{E}}_{Coul}=-\frac{2\Omega^2}{\lambda^2}.
\ee
The generators of hidden symmetries also transform one into the other on the energy surface
\be
\mathcal{M}_{(a)osc}=\left(\imath\lambda\sqrt[4]{2{\cal \tilde{I}}}\right)^{n_a}\mathcal{M}_{(a)Coul}
\ee
Finally, let us write down the relation between generators of conformal symmetries defined on "tilded" and untilded spaces.
\be
\mathcal{H}_0= \lambda \mathcal{\tilde{H}}_0\sqrt{ 2\mathcal{\tilde{K}}}, \qquad \mathcal{D}=2\mathcal{\tilde{D}},
\qquad \mathcal{K}= \frac{2\sqrt{2\mathcal{\tilde{K}}}}{\lambda}.
\ee

In this Section we  transformed   deformed oscillator into deformed Coulomb problem,  preserving intact angular coordinates.
Performing proper transformations of angular part of oscillator, including its reduction, we can get variety of superintegrable deformations of Coulomb problem. However, they will  belong to the same class of systems  under consideration, since the latter are formulated in most general, action-angle variables, terms.

\section{Spherical and pseudospherical generalizations}
Oscillator and Coulomb systems  admit superintegrable generalizations to $N$-dimensional spheres and two-sheet hyperboloids (pseudospheres),
 which are given by the Hamiltonians \cite{higgs}
\be
\label{Hv}
\mathbb{S}^{N}:
\qquad {\cal H}_V=\frac{p_{\chi}^2}{2 r_0^2}+\frac{{\cal I}}{ r_0^2 \sin^2\chi}+V(\tan\chi),
\qquad
\mathbb{H}^{N}:\quad {\cal H}_V=\frac{p_{\chi}^2}{2 r_0^2}+\frac{{\cal I}}{ r_0^2\sinh^2\chi}+V(\tanh\chi)%\quad \{p_\chi, \chi\}=1
%\quad\{I_i,\Phi^0_i\}=\delta_{ij}, \Phi^0_i\in [0, 2\pi),\;i,j=1,\ldots,N-1,
\ee
with  the  potentials
\begin{align}
\mathbb{S}^{N}:
&\qquad  V_{osc}(\tan\chi)=\frac{r^2_0\omega^2\tan^2\chi}{2},
&& V_{Coul}(\tan\chi)=-\frac{\gamma}{r_0}\cot\chi,
\label{Vsph}
\\
\mathbb{H}^{N}:
&  \qquad  V_{osc}(\tanh\chi)=\frac{r^2_0\omega^2\tanh^2\chi}{2},
&&  V_{Coul}(\tanh\chi)=-\frac{\gamma}{r_0}\coth\chi.
\label{Vhyp}
\end{align}
 Here ${\cal I}$ is a quadratic Casimir element of the orthogonal algebra $so(N)$.  To get  integrable deformations of these systems,  we  replace it, as in Euclidean case,  by some
integrable (angular) Hamiltonian  depending on the action variables \cite{hlnsy}.
The  particular angular Hamiltonian \eqref{angular0} defines superintegrable systems
as in the flat case.
About decade ago the so-called $\kappa$-dependent formalism was developed \cite{kappa} where the oscillator and Coulomb  systems on plane and on the
  two-dimensional
sphere and  hyperboloid  were described in the unified way.

Introduce, following that papers,
\be
\label{T}
T_{\kappa}=\frac{S_{\kappa}}{C_{\kappa}}
\qquad\text{with}\qquad
C_{\kappa}(x)=
\begin{cases}
\cos{\sqrt{\kappa}x}  &   \kappa>0,  \\
 1  & \kappa=0,\\
 \cosh{\sqrt{-\kappa}x}  &  \kappa<0,
\end{cases}
\qquad\quad
S_{\kappa}(x)=
\begin{cases}
\displaystyle
\frac{\sin{\sqrt{\kappa}x}}{{\sqrt{\kappa}}}   & \kappa>0,  \\
 x  & \kappa=0,\\
 \displaystyle
 \frac{\sinh{\sqrt{-\kappa}x}}{\sqrt{-\kappa}}  & \kappa<0,
\end{cases}
\ee
where the parameter $\kappa$  in two-dimensional  case
coincides with the curvature of (pseudo)sphere,
\be
\label{kappa}
\mathbb{S}^{N}: \quad\kappa=\frac{1}{r_0^2},
\qquad\qquad
\mathbb{H}^{N}: \quad\kappa=-\frac{1}{r_0^2}.
\ee
The case  $\kappa=\pm 1$ corresponds to a unit sphere/pseudosphere.
 For $\kappa\neq 0$ we  identify
 \be
 \label{x-kappa}
 x=r_0\chi=\frac{\chi}{\sqrt{\kappa}},
 \qquad
  p_x= \frac{p_\chi}{r_0} = \sqrt{\kappa} p_\chi.
  \ee
The  "holomorphic factorization" approach to two-dimensional systems  was combined with $\kappa$-dependent formalism by Ranada.
Let  us show that it can be straightly extended to any dimension.
For this purpose introduce   a (pseudo)spherical analog of  $z$, $\bar z$ coordinates  and obtain their Poisson bracket:
\be
\label{z-kappa}
z=\sqrt{|\kappa|}\frac{{p_\chi}}{\sqrt2}+\frac{\imath\sqrt{\cal I}}{T_\kappa},
\qquad
\{\bar{z},z\}=\frac{\imath(z-\bar{z})^2}{2\sqrt{2\cal I}}-\imath\kappa {\sqrt{{2\cal I}}}.
\ee
The Poisson brackets between $z$,  $u_a$ and ${\bar u}_a$ remain unchanged [see relations \eqref{z-z}].

In these  terms the $\kappa$-deformed Hamiltonian reads
\be
\label{Hkap}
\mathcal{H}_{osc/Coul}=\mathcal{H}_0+V_{osc/Coul},
\qquad
\mathcal{H}_{0}=\frac{p_{r}^2}{2}+\frac{{\cal I}}{S_{\kappa}^2}+\kappa  \mathcal{ I}
= z \bz+\kappa \mathcal{I},
\ee
where using \eqref{T}, \eqref{kappa},
\eqref{x-kappa}, \eqref{z-kappa},   the oscillator and Coulomb potentials on sphere \eqref{Vsph} can be expressed
as follows:
\be
\label{Vkap}
V_{osc}=\frac{\omega^2T^2_{\kappa}}{2}=-\frac{{2\omega^2{\cal I}}}{ ({\bar z}- { z})^2} ,
\qquad
V_{Coul}=-\frac{\gamma}{T_{\kappa}}=-\imath\gamma\frac{\bz-z}{2\sqrt{\mathcal{I}}}.
\ee
The (local  and global) constants of motion and related quantities have the same expressions  in terms of  $z$, $\bar z$ as in the flat case, with the Hamiltonians shifted  in agreement with \eqref{Hkap}
\be
\label{shift}
{\cal H}\to {\cal H}-\kappa{\cal I}.
\ee
%
%\subsubsection*{Free particle}

For the free system on sphere, ${\cal H}_0$, the most of  Poisson brackets among the integrals
survive from the flat case [see relations \eqref{M-M}, \eqref{M-bMcal} and \eqref{M-Mcal}]. The only brackets, which acquire
 extra $\kappa$-dependent terms, are:
\begin{gather}
\label{M-Mkap}
\{M_a,\overline{M}_b  \}
=\left(\frac{ \imath\kappa \sqrt{ 2{\cal I}} } { {\cal H}_0-\kappa{\cal I} }
-\frac{\imath\delta_{ab}}{k_a^2I_a}\right)
M_a\overline{M}_b
=-\frac{\imath\delta_{ab}}{k_a^2} I_a^{\frac{1}{k_a}-1}({\cal H}_0-\kappa{\cal I})
+\frac{ \imath\kappa \sqrt{ 2{\cal I}} } { {\cal H}_0-\kappa{\cal I} } M_a\overline{M}_b,
\\
\{{\cal M}_a,\overline{{\cal M}}_b  \}
=\imath\left(\frac{ \kappa n_an_b\sqrt{ 2{\cal I}} } { {\cal H}_0-\kappa{\cal I} }
-\frac{ m_a^2\delta_{ab}}{I_a}\right){\cal M}_a\overline{{\cal M}}_b.
\end{gather}
%Note that apart from the Hamiltonian shift \eqref{shift},  a linear correction on $\kappa$ appears in above equations,
%\begin{align}
%&\{{\cal M}_a,\overline{\mathcal{ M}}_b  \}=\imath{\cal M}_a\overline{\mathcal{ M}}_b
%m_a m_b \Big(\kappa \frac{n_an_b\sqrt{2{ \cal I}}}{\mathcal{H}_{\kappa}-\kappa { \cal I}}-\frac{\delta_{ab}}{u_a\bar{u_b}} \Big)
%\label{inconfalgkappa}
%\end{align}

Let us write down also the  deformation of conformal algebra \eqref{ca}
\be
\{\mathcal{H}_0,\mathcal{D}\}=2(\mathcal{H}_0-\kappa { \cal I})(1+2\kappa \mathcal{K}) ,
\qquad
\{\mathcal{H}_0, \mathcal{K} \}=\mathcal{D}(1+2\kappa \mathcal{K}) ,
\qquad
\{\mathcal{D},\mathcal{K}\}=2\mathcal{K}(1+2\kappa \mathcal{K} ).
\ee
%\subsubsection*{Coulomb problem}

\smallskip

For the Coulomb problem on sphere, the Poisson brackets between the local integrals \eqref{I-Mcoul} remain unaffected, while
the relations \eqref{M-Mcoul} undergo a similar modification:
\be
\label{M-Mcoulkap}
\begin{aligned}
\big\{ M_a^{Coul},\overline{M}_b^{Coul}  \big\}
&=\left[\frac{\imath\sqrt{2{\cal I}}\left(\frac{\gamma^2}{4{\cal I}^2}+\kappa\right)}
    {{\cal H}_{coul}-\kappa{\cal I}+\frac{\gamma^2}{4{\cal I}^2}}
    -\frac{\imath\delta_{ab}}{k_a^2I_a}\right] M_a^{Coul}\overline{M}_b^{Coul}
\\
&=
\imath\sqrt{2{\cal I}}\left(\frac{\gamma^2}{4{\cal I}^2}+\kappa\right)\frac{ M_a^{Coul}\overline{M}_b^{Coul} }
    {{\cal H}_{coul}-\kappa{\cal I}+\frac{\gamma^2}{4{\cal I}^2}}
- \frac{\imath\delta_{ab}}{k_a^2}I_a^{\frac{1}{k_a}-1}\left({\cal H}_{Coul}-\kappa{\cal I}+\frac{\gamma^2}{4{\cal I}^2}\right).
\end{aligned}
\ee
%\subsubsection*{Oscillator}

\smallskip

Consider now the spherical system \eqref{Hv} with the oscillator potential.
Line for the flat case, the integrals of motion are based on  the simpler local
quantities $A$ and $B$,
\be
\label{Aa-sp}
A_a=(z+\frac{\imath\omega T_\kappa}{\sqrt2})u_a^{\frac1{k_a}},
\qquad
B_a=(z-\frac{\imath\omega T_\kappa}{\sqrt2})u_a^{\frac1{k_a}},
\qquad
 {\cal M}^{osc}_a=(A_aB_a)^{n_a},
\ee
which evolve in time under  the following rule:
\be
\label{H-A-sp}
\{{\cal H}_{osc}, A_a\}=\imath\omega(1+\kappa T_\kappa^2)  A_a,
\qquad
\{{\cal H}_{osc}, B_a\}=-\imath\omega(1+\kappa T_\kappa^2)   B_a.
\ee
They are $\kappa$-deformations of the harmonic oscillating quantities
\eqref{Aa}, \eqref{H-A} in the flat case. Unlike them, they do not oscillate
harmonically, but
the product $A_a B_b$ is still preserved.

The Poisson brackets between local quantities can be  calculated explicitly
giving rise to $\kappa$-deformations of the relations \eqref{A-barB},
\eqref{A-barA}, \eqref{B-barB}:
\begin{gather}
\label{A-Bkap}
\{A_a, B_b\}=-\frac{\imath \kappa\omega  T_\kappa^2}{z^2 + \frac{\omega ^2T_\kappa^2}{2}}A_a  B_b,
\qquad
\{A_a,\bar B_b\}=-\frac{\imath \delta_{ab}}{k_a^2 I_a}A_a\bar B_a
+\frac{ \imath\kappa  \sqrt{ 2{\cal I}}  A_a\bar A_b} { {\cal H}_{osc}-\kappa{\cal I} +\omega  \sqrt{2{\cal I}}},
\\
\label{A-barAkap}
\{A_a,\bar A_b\}=\imath
\frac{\kappa(\sqrt{2{\cal I}} -2\omega T_\kappa) -2\omega  }{\mathcal{H}_{osc}-\kappa {\cal I}+\omega \sqrt{2{\cal I}}}A_a\bar A_b
-\frac{\imath\delta_{ab}}{k_a^2}I_a^{\frac{1}{k_a}-1}
(\mathcal{H}_{osc}-\kappa {\cal I}+\omega \sqrt{2{\cal I}}),
\\
\label{B-barBkap}
\{B_a,\bar B_b\}=\imath
\frac{\kappa(\sqrt{2{\cal I}} +2\omega T_\kappa) +2\omega  }{\mathcal{H}_{osc}-\kappa {\cal I}-\omega \sqrt{2{\cal I}}}A_a\bar A_b
-\frac{\imath\delta_{ab}}{k_a^2}I_a^{\frac{1}{k_a}-1}
(\mathcal{H}_{osc}-\kappa {\cal I}-\omega \sqrt{2{\cal I}}).
\end{gather}
The Poisson brackets between the true integrals of motion ${\cal M}^{osc}_a$, ${\cal M}^{Coul}_a$ and their conjugate
are based on the local brackets
\eqref{M-Mcoulkap}, \eqref{A-Bkap}, \eqref{A-barAkap}, \eqref{B-barBkap}
and can be easily obtained.

\section{Example of spherical part of  higher-order superintegrable system with separation of variables}
In previous Sections we extended "holomorphic factorization approach" to higher-dimensional superintegrable systems with oscillator and Coulomb potentials, including those on spheres and hyperboloids.
For this purpose we separated the "radial" and "angular" variables in these systems. Then we combined the radial coordinate and momentum in single complex coordinate parameterizing Klein model of Lobachevsky space, and combined  "angular" coordinates  and their conjugated momenta in complex coordinates
 by the use of action-angle variables.
However, action-angle variables are not in common use  in present math-physical society, and their explicit expressions are  not common even for the
such  textbook models like oscillator and Coulomb problems.

For clarifying  the relation of the above formulations of constants of motion with their conventional representations  first present the
action-angle variables of the angular part(s) of non-deformed, oscillator and Coulomb systems (on Euclidean space, sphere and hyperboloids).
 Its  Hamiltonian is given by  the quadratic Casimir element of $so(N)$ algebra on $(N-1)$-sphere, ${\cal I}=L_{N}^2/2$. It can be decomposed by the eigenvalues of the embedded $SO(a)$ angular momenta  defining the action variables
 $I_a $. For the details of derivation  of  their explicit expressions, for those of  conjugated angle variables we refer to   Appendix in Ref.~\cite{hlnsy}.
  The action variables are given by the expressions
  \be
  I_a=\sqrt{j_{a+1}}-\sqrt{j_{a}},\quad {\rm where}\quad j_{a+1}=p^{2}_{a}+\frac{j_{a}}{\sin^2 \theta_{a}},\qquad j_0=0,\qquad a=1,\ldots N-1.
  \ee
 This gives rise   the angular Hamiltonian which belongs to the family \eqref{angular0}
\be
{\cal I}=\frac12\left(\sum_{a=1}^{N-1}I_a \right)^2.
\ee
Its substitution to the Hamiltonians \eqref{flat},\eqref{Hv} leads to well-known oscillator and Coulomb systems on the Euclidean spaces, spheres and hyperboloids.

The expressions for angle variables are more complicated,
\be\Phi_a=\sum_{l=a}^{N-1}a_l+ \sum_{l=a+1}^{N-1}b_l,\qquad{\rm where }\qquad
a_l=\arcsin\sqrt{\frac{j_{l+1}}{j_{l+1}-j_{l}}} \cos{\theta_l},\quad b_l=\arctan \frac{\sqrt{j_l}\cos\theta_l}{p_l\sin\theta_l}.%,
\ee%\quad
%\sqrt{j_a}=\sum_{m=1}^{a}I_m
Direct transformations give the following
expressions for $u_a $  coordinates:
\be
u_a=\sqrt{\sqrt{j_{a+1}}-\sqrt{j_a}}\;{\rm e}^{\imath a_a} \prod_{l=a+1}^{N-1}{\rm e}^{\imath(a_l+b_l)},
\ee
with
\be
{\rm e}^{\imath a_l}=
\frac{p_l\sin\theta_l +\imath\sqrt{j_{l+1}}\cos\theta_l} {\sqrt{j_{l+1}-j_{l}}  },\qquad {\rm e}^{\imath b_l}=\frac{p_l\sin\theta_l +\imath\sqrt{j_{l}}\cos\theta_l} {\sqrt{j_{l+1}-j_{l}} \sin\theta_l }
\ee
With these expressions at hand we can express  ``holomorphic representation" of constants of motion via initial coordinates.
In two-dimensional  case it has transparent  relation with conventional representations of hidden constants of motion, like Fradkin tensor (for the oscillator) and Runger-Lenz vector (for Coulomb problem) \cite{PANShmavon}. In the higher dimensional cases the relation of these two representations is more complicated.

\smallskip

This construction could easily be modified to the system whose Hamiltonian is given in the angle variables by the
 generic expression \eqref{angular0}. We define it
by the recurrence relation
\be
{\cal I}\equiv \frac12 {j_N}, \qquad j_{a}=p^2_{a-1}+\frac{j_{a-1}}{\sin^2 k_{a-1}\theta_{a-1}},\qquad a=1,\ldots N-1, \qquad j_0=c_0.
\ee
It describes  particle moving on the space (spherical segment)   equipped with the diagonal metric
\be
ds^2=g_{ll}(d\theta_l)^2, \qquad g_{N-1.N-1}=1,\qquad  g_{ll}=\prod_{m=l}^{N-1}\sin^2 k_{m}\theta_{m}
\ee
and interacting with the  potential field
\be
U=\frac{c_0}
{ \prod_{l=1}^{N-1} \sin^2 k_l \theta_l}.
\ee
Redefining the angles, $\theta_a\to \theta_{a}/k_a$, we can represent the above metric  in the form
\be
ds^2=\frac{1}{k_{a}^2}\prod_{a=1}^{N-1}\sin^2\theta_{a}(d\theta_a)^2.
\ee

It is obvious, that  the functions $j_k(\theta_а, p_а   )$ define commuting constants of motions of the system.
Similar to derivation given in Appendix of Ref.~\cite{hlnsy} we can use  action-angle variable formulation, and find that the Hamiltonian is given by the expression \eqref{angular0}. The
 action variables are related  with the initial ones by the expressions
\be
{I}_a=\frac{1}{2\pi}\int^{\theta_{min}}_{\theta_{max}}\sqrt{j_{a+1}-\frac{j_{a}}{\sin^2k_a \theta_a}}\,d\theta_a =\frac{\sqrt{j_{a+1}}-\sqrt{j_{a}} }{k_a}  \qquad\Rightarrow\qquad j_a=\Bigg( \sum^{N-1}_{a=1}k_a I_a+ c_0\Bigg)^2.
  \ee
  The angle variables read
  \be
  \Phi_a=\sum^{N-1}_{l=a} \frac{k_a}{k_l} a_l + \sum_{l=a+1}^{N-1} \frac{k_a}{k_l}b_l,\qquad
  a_l=\arcsin \sqrt{\frac{j_{l+1}}{j_{l+1}-j_{l}}}\cos{k_l\theta_l},
  \qquad  b_l=\arctan\frac{\sqrt{j_l}\cos\theta_l}{p_l\sin k_l\theta_l}.
  \ee
Thus,
\be
u_a= \frac{1}{k_a}\sqrt{\sqrt{j_{a+1}}-\sqrt{j_a}}  \prod_{l=a}^{N-1} \left(\frac{p_l\sin k_l\theta_l +\imath\sqrt{j_{l+1}}\cos k_l\theta_l} {\sqrt{j_{l+1}-j_{l}}  }\right)^{\frac{k_a}{k_l}}
\prod_{l=a+1}^{N-1}\left(\frac{p_l\sin k_l \theta_l +\imath\sqrt{j_{l}}\cos k_l \theta_l} {\sqrt{j_{l+1}-j_{l}} \sin\theta_l }\right)^{\frac{k_a}{k_l}}.
 \ee
Hence, we constructed the superintegrable system with higher order constants of motion, which admits separation of variables.
Since the classical spectrum of its angular part is isospectral with the   "angular  Calogero model", we can state that they become, under appropriate choice of constants $k_i$, $c_0$,  canonically equivalent with angular part of rational Calogero model \cite{flp}.
In fact this means equivalence of these two systems.  However, we can't present explicit mapping of one system to other.
\section{Concluding remarks}
In this work we investigated  superintegrable deformations of oscillator and Coulomb problems separating their "radial" and "angular" parts, where the latter was described in terms of action-angle variables. We encoded phase space coordinates in the complex ones: the complex coordinate $z$ involved radial variables parameterizing Klein model of Lobachevsky plane, and complex coordinates $u_a$ encoding  action-angle variables of the angular part.
Then we combined the whole set of constants of motion (independent from Hamiltonian)  in  $N-1$ holomorphic  functions $\mathcal{M}_a$, generalizing the so-called "Holomorphic factorization" earlier developed for two-dimensional generalized oscillator/Coulomb systems. Then we presented their algebra, which among nontrivial relations possesses chirality property $\{\mathcal{M}_a, \mathcal{M}_a\}=0$. Hence, presented representation can obviously considered as a classical trace of ''quantum factorization" of respective Hamiltoinian. Seems that  it could be used for the construction of  supersymmetric extensions of these systems. The lack of given representation is the use of  the action-angle formulation of the angular parts of the original systems.

In this context  one should mention the earlier work \cite{hkl}, where symmetries of the angular parts of conformal mechanics (and those with additional oscillator potential) were related with the symmetries of the whole system by the use of  coordinate $z$ and conformal algebra generators \eqref{uk2}.
That study was done in most general terms, without referring to action-angle variables and to specific form of angular part. Quantum mechanical aspects were also considered there.
Hence, it seems  to be natural to combine these two approaches for and at first,  exclude the action-angle argument from present formulations, and at second, use presented constructions for the quantum considerations of systems, in particular, for construction of spectrum and wavefunctions within operator approach. We are planning to present this elsewhere.

\acknowledgments
This work was  partially supported by the Armenian
State Committee of Science Grants No. 15RF-039 and No. 15T-1C367 and by Grant No. mathph-4220 of the
Armenian National Science and Education Fund based in New York (ANSEF). It was done within ICTP programs NET68 and OEA-AC-100 and
within program of Regional Training Network on Theoretical Physics
sponsored by Volkswagenstiftung Contract nr. 86 260.


\begin{thebibliography}{99}

\bibitem{calogero}
F.~Calogero,
\emph{Solution of a three-body problem in one-dimension},
\href{http://dx.doi.org/10.1063/1.1664820}{J. Math. Phys. {\bf 10} (1969) 2191};
\emph{Solution of the one-dimensional N-body problems with quadratic and/or inversely quadratic
pair potentials},
\href{http://dx.doi.org/10.1063/1.1665604}{{\sl ibid.} {\bf 12} (1971) 419};
J. Moser,
\emph{Three integrable Hamiltonian systems connected with isospectral deformations},
\href{http://dx.doi.org/10.1016/0001-8708(75)90151-6}
{Adv. Math. {\bf 16} (1975) 197}.

\bibitem{algebra}
M.~Olshanetsky and  A.~Perelomov,
 \emph{Classical integrable finite dimensional systems related to Lie algebras},
\href{http://dx.doi.org/10.1016/0370-1573(81)90023-5}
{Phys. Rept.  {\bf 71} (1981) 313};
\emph{Quantum integrable systems related to Lie algebras},
\href{http://dx.doi.org/10.1016/0370-1573(83)90018-2 }
{{\sl ibid.}  {\bf 94} (1983) 313}.

\bibitem{woj83}
 S.~Wojciechowski,
 \emph{Superintegrability of the Calogero-Moser system},
 \href{http://dx.doi.org/10.1016/0375-9601(83)90018-X}
 {Phys. Lett. A {\bf 95} (1983) 279}.

\bibitem{fubini}
 V.~de Alfaro, S.~Fubini and G.~Furlan, \emph{Conformal Invariance In Quantum Mechanics},
  {\it Nuovo Cim.}  A  {\bf 34} (1976) 569.

\bibitem{higgs}
P.W. Higgs,
\emph{Dynamical symmetries in a spherical geometry. 1},
\href{http://dx.doi.org/10.1088/0305-4470/12/3/006}{J. Phys. A {\bf 12} (1979) 309};
H.I. Leemon,
\emph{Dynamical symmetries in a spherical geometry. 2},
\href{http://dx.doi.org/10.1088/0305-4470/12/4/009}{J. Phys. A {\bf 12} (1979) 489}.

\bibitem{hlnsy}
 T.~Hakobyan, O.~Lechtenfeld, A.~Nersessian, A.~Saghatelian and V.~Yeghikyan,
 \emph{Integrable generalizations of oscillator and Coulomb systems via action-angle variables},
 Phys. Lett. A {\bf 376}, 679 (2012),
 \href{http://arxiv.org/abs/1108.5189}{arXiv:1108.5189}.

\bibitem{rapid}
  T.~Hakobyan, O.~Lechtenfeld and A.~Nersessian,
  \emph{Superintegrability of generalized Calogero models with oscillator or Coulomb potential,}
  Phys. Rev. D {\bf 90}, no. 10, 101701 (2014)
 \href{http://arxiv.org/abs/1409.8288}{arXiv:1409.8288 }.

\bibitem{flp}
M.~Feigin, O.~ Lechtenfeld, A.~Polychronakos,
\emph{The quantum angular Calogero-Moser model},
JHEP {\bf 1307} (2013) 162,
\href{http://arxiv.org/abs/1305.5841}{arXiv:1305.5841}.

\bibitem{runge}
T.~Hakobyan and A.~Nersessian,
\emph{Runge-Lenz vector in Calogero-Coulomb problem},
 Phys. Rev. A \textbf{92} (2015) 022111
\href{http://arxiv.org/abs/1504.00760}{arXiv:1504.00760}.
\bibitem{francisco}
  F.~Correa, T.~Hakobyan, O.~Lechtenfeld and A.~Nersessian,
  \emph{Spherical Calogero model with oscillator/Coulomb potential: quantum case},
  Phys. Rev. D {\bf 93} (2016)  125009,
  \href{http://arxiv.org/abs/1604.00027}{arXiv:1604.00027 };
  \emph{Spherical Calogero model with oscillator/Coulomb potential: classical case},
  Phys. Rev. D {\bf 93} (2016)  125008
 \href{http://arxiv.org/abs/1604.00026}{arXiv:1604.00026 }.

\bibitem{Calogero-Stark}
T. Hakobyan and A. Nersessian,
\emph{Integrability and separation of variables in Calogero-Coulomb-Stark and two-center
Calogero-Coulomb systems},
\href{http://arxiv.org/abs/1509.01077}
{arXiv:1509.01077}.

\bibitem{TTW}
F. Tremblay, A.V. Turbiner, and P. Winternitz,
\emph{An infinite family of solvable and integrable quantum systems on a plane},
J. Phys. A {\bf 42} (2009) 242001,
\href{http://arxiv.org/abs/0904.0738}
{arXiv:0904.0738}.

\bibitem{pw}
S.~Post and P.~Winternitz,
\emph{An infinite family of superintegrable deformations of the Coulomb potential},
J. Phys. A {\bf 43}, 222001 (2010),
\href{http://arxiv.org/abs/1003.5230}{arXiv:1003.5230}.

\bibitem{kalnins}
E.G. Kalnins,  K.M. Kress, and W. Miller,
\emph{Superintegrability and higher order constants for quantum
systems},
J. Phys. A {\bf 43} (2010) 265205,
\href{http://arxiv.org/abs/1002.2665}{arXiv:1002.2665}.


\bibitem{lny}
O.~Lechtenfeld, A.~Nersessian and V.~Yeghikyan,
\emph{Action-angle variables for dihedral systems on the circle},
Phys. Lett. A {\bf 374}, 4647 (2010),
\href{http://arxiv.org/abs/1005.0464}
{arXiv:1005.0464}.

\bibitem{ranada}
M.~F.~Ranada,
\emph{The Tremblay-Turbiner-Winternitz system on spherical and hyperbolic spaces:
Superintegrability, curvature-dependent formalism and complex factorization},
  J. Phys. A {\bf 47} (2014) 165203,
  \href{http://arxiv.org/abs/1403.6266}{arXiv:1403.6266};
\emph{A new approach to the higher order superintegrability of the Tremblay-Turbiner-Winternitz system},
  J. Phys. A {\bf 45} (2012) 465203,
  \href{https://arxiv.org/abs/1211.2919}{arXiv:1211.2919};
\emph{Higher order superintegrability of separable potentials with a new approach to the Post-Winternitz system},
  \href{http://dx.doi.org/10.1088/1751-8113/46/12/125206}{J. Phys. A {\bf 46} (2013) 125206}.

\bibitem{PANShmavon}
T. Hakobyan, A. Nersessian, and H. Shmavonyan,
\emph{Lobachevsky geometry in TTW and PW systems},
\href{http://arxiv.org/abs/1512.07489}{arXiv:1512.07489}.
\bibitem{lobach}
  T.~Hakobyan and A.~Nersessian,
 \emph{Lobachevsky geometry of (super)conformal mechanics},
  Phys. Lett. A {\bf 373} (2009) 1001,
   \href{http://arxiv.org/abs/0803.1293}{arXiv:0803.1293}.

 C.~Burdik and A.~Nersessian,
 \emph{Remarks on multi-dimensional conformal mechanics},
  SIGMA {\bf 5} (2009) 004,
 \href{http://arxiv.org/abs/0901.1644}{arXiv:0901.1644}.

\bibitem{sphCal}
T.~Hakobyan, A.~Nersessian and V.~Yeghikyan,
\emph{Cuboctahedric Higgs oscillator from the Calogero model}
J.  Phys. A  {\bf 42}  (2009) 205206,
\href{http://arxiv.org/abs/0808.0430}{arXiv:0808.0430};
T.~Hakobyan, S.~Krivonos, O.~Lechtenfeld and A.~Nersessian,
\emph{Hidden symmetries of integrable conformal mechanical systems},
Phys. Lett.  A {\bf 374} (2010) 801,
\href{http://arxiv.org/abs/0908.3290}{arXiv:908.3290};
T.~Hakobyan, O.~Lechtenfeld, A.~Nersessian and A.~Saghatelian,
\emph{Invariants of the spherical sector in conformal mechanics}
J. Phys. A  {\bf 44} (2011) 05520,
\href{http://arxiv.org/abs/1008.2912}{arXiv:1008.2912};
T.~Hakobyan, O.~Lechtenfeld, and A.~Nersessian,
\emph{The spherical sector of the Calogero model as a reduced matrix model},
Nucl. Phys. B  {\bf 858} (2012) 250,
\href{http://arxiv.org/abs/1110.5352}{arXiv:1110.5352};
  F.~Correa and O.~Lechtenfeld,
  \emph{The tetrahexahedric angular Calogero model},
  JHEP {\bf 1510} (2015) 191,
  \href{http://arxiv.org/abs/1508.04925}{arXiv:1508.04925}.


\bibitem{GalajBH}
 S.~Bellucci, A.~Galajinsky, E.~Ivanov and S.~Krivonos,
 \emph{AdS(2)/CFT(1), canonical transformations and superconformal mechanics,}
  Phys. Lett. B {\bf 555} (2003) 99
 % doi:10.1016/S0370-2693(03)00040-6
 \href{http://arxiv.org/abs/hep-th/0212204}{hep-th/0212204};
  A.~Galajinsky,
 \emph{Particle dynamics near extreme Kerr throat and supersymmetry},
  JHEP {\bf 1011} (2010) 126
 \href{http://arxiv.org/abs/1009.2341}{arXiv:1009.2341};
 A.~Galajinsky and K.~Orekhov,
  \emph{N=2 superparticle near horizon of extreme Kerr-Newman-AdS-dS black hole,}
  Nucl. Phys. B {\bf 850} (2011) 339
 % doi:10.1016/j.nuclphysb.2011.04.015
   \href{http://arxiv.org/abs/1103.1047}{arXiv:1103.1047};
A. Galajinsky, A. Nersessian, and A. Saghatelian,
\emph{Superintegrable models related to near horizon extremal Myers-Perry black hole in arbitrary dimension},
JHEP {\bf 2013} (2013) 2,
\href{http://arxiv.org/abs/1303.4901}{arXiv:1303.4901};
 A.~Galajinsky and K.~Orekhov,
 \emph{On the near horizon rotating black hole geometries with NUT charges,}
  Eur.\ Phys.  J. C {\bf 76} (2016)  477,
   \href{http://arxiv.org/abs/1604.08056}{arXiv:1604.08056}.

\bibitem{ter}
V.~Ter-Antonian, \emph{Dyon oscillator duality},
  \href{http://arxiv.org/abs/quant-ph/0003106}{quant-ph/0003106};
  A.~Nersessian and V.~M.~Ter-Antonian,
\emph{Anyons, monopole and Coulomb problem},
  Phys. Atom. Nucl.  {\bf 61} (1998) 1756,
   Yad. Fiz.  {\bf 61} (1998) 1868,
   \href{http://arxiv.org/abs/physics/9712027}{physics/9712027}.


\bibitem{ntt}
  A.~Nersessian, V.~Ter-Antonian, and M.~M.~Tsulaia,
\emph{A Note on quantum Bohlin transformation},
  Mod. Phys. Lett. A {\bf 11} (1996) 1605,
\href{http://arxiv.org/abs/hep-th/9604197}{hep-th/9604197};
 A.~Nersessian and V.~Ter-Antonian,
 \emph{'Charge dyon' system as the reduced oscillator},
  Mod.\ Phys. Lett. A {\bf 9} (1994) 2431
   \href{http://arxiv.org/abs/hep-th/9406130}{hep-th/9406130};
  L.~G.~Mardoyan, A.~N.~Sisakian and V.~M.~Ter-Antonian,
\emph{Hidden symmetry of the Yang-Coulomb system},
  Mod. Phys. Lett. A {\bf 14} (1999) 1303,
   \href{http://arxiv.org/abs/hep-th/9803010}{hep-th/9803010}.

\bibitem{kappa}
 P. Dombrowski and  J.Zitterbarth,
 \emph{On the planetary motion in the 3-Dim standard spaces of constant curvatur},
 Demonstratio Mathematica {\bf 24} (1991) 375;
 A. Ballesteros, F.J.~Herranz ,  M.A.~del Olmo, and  M.~Santander,
 \emph{Quantum structure of the motion groups of the two-dimensional Cayley-Klein geometries},
\href{http://dx.doi.org/10.1088/0305-4470/26/21/019}{Phys. A  {\bf  26} (1993)  5801};
  M.F.~Ra\~nada and  M.~Santander,
\emph{Superintegrable systems on the two-dimensional sphere S2 and the hyperbolic plane H2},
\href{http://dx.doi.org/10.1063/1.533014J}{Math. Phys. {\bf  40} (1999) 5026}.

\bibitem{hkl}
T.~Hakobyan, D.~Karakhanyan, and O.~Lechtenfeld,
\emph{The structure of invariants in conformal mechanics},
 Nucl. Phys. B {\bf 886} (2014) 399,
\href{http://arxiv.org/abs/1402.2288}{arXiv:1402.2288}.

\end{thebibliography}
\end{document}